%% file: IDM.tex
\documentclass[sigconf]{acmart}

\usepackage{booktabs} 
\usepackage{subfigure}

\setcopyright{rightsretained}

\copyrightyear{2017}
\acmYear{2017}
\setcopyright{rightsretained}
\acmConference{IDM Workshop @ CIKM'17}{November, 2017}{Singapore}
\acmDOI{http://dx.doi.org/10.1145/xxx}

\begin{document}

\title[Interpreting Contextual Effects By Contextual Modeling]{Interpreting Contextual Effects By Contextual Modeling In Recommender Systems}

\author{Yong Zheng}
\affiliation{%
  \institution{School of Applied Technology}
  \streetaddress{Illinois Institute of Technology}
  \city{Chicago}
  \state{IL}
  \postcode{60616, USA}
}
\email{yzheng66@iit.edu}


\begin{abstract}
Recommender systems have been widely applied to assist user's decision making by providing a list of personalized item recommendations. Context-aware recommender systems (CARS) additionally take context information into considering in the recommendation process, since user's tastes on the items may vary from contexts to contexts. Several context-aware recommendation algorithms have been proposed and developed to improve the quality of recommendations. However, there are limited research which explore and discuss the capability of interpreting the contextual effects by the recommendation models. In this paper, we specifically focus on different contextual modeling approaches, reshape the structure of the models, and exploit how to utilize the existing contextual modeling to interpret the contextual effects in the recommender systems. We compare the explanations of contextual effects, as well as the recommendation performance over two-real world data sets in order to examine the quality of interpretations.
\end{abstract}

%
%




\maketitle

\input{samplebody-conf}

\bibliographystyle{ACM-Reference-Format}
\bibliography{sigproc}

\end{document}

%% file: samplebody-conf.tex
\section{Introduction}
Recommender systems provide personalized suggestions of products to end-users in a variety of settings. It has been applied to several domains, such as e-commerce (e.g., Amazon and eBay), online streaming (e.g., Netflix and Pandora), social media (e.g., Facebook and Twitter), and so forth.

One way to view the recommendation problem by the collaborative recommendation approach is to consider the users and items as forming a matrix $U \times I$ where the entries in the matrix are known ratings by particular users for given items. In this framework, collaborative recommendation becomes the task of creating a function for predicting the likely values of unknown cells in this matrix. i.e.,  \textit{R}: \textit{Users} $\times$ \textit{Items} $\rightarrow$ \textit{Ratings}.

In recent years, context-awareness in the recommender systems raised the research attentions. The list of the recommendations cannot stand alone without considering contexts, since user's tastes may vary from contexts to contexts. For example, a user may choose a different movie if he is going to watch the movie with kids rather than with his partner. A user may prefer a fast food restaurant for quick lunch by himself, while he may choose a formal restaurant if he is going to have a business dinner with the colleagues. As a result, context-aware recommendation turns the prediction task into a \textit{multidimensional} rating function -- \textit{R}: \textit{Users} $\times$ \textit{Items} $\times$ \textit{Contexts} $\rightarrow$ \textit{Ratings}~\cite{carsmobasher}.

Several context-aware recommendation algorithms were proposed and developed in the past decade. They explore different ways to incorporate context information (such as time, location, weather, companion, etc)~\cite{zheng2015revisit} into the recommendation models, in order to improve the quality of item recommendations. Among these effective algorithms, many of them are machine learning based approaches which are able to significantly improve the recommendations but hard to be interpreted, such as the models based matrix factorization. Therefore, there are limited research that try to utilize the model to interpret the contextual effects in the recommender systems. It is not that easy to understand how context information take effects in the recommendation process, and how can they affect the quality of the recommendations.

In this paper, we specifically focus on different contextual modeling approaches, and make the first attempt to reshape the structure of the models, in order to further exploit how to utilize the existing contextual modeling to interpret the contextual effects in the recommender systems.

\section{Related Work}
In this section, we introduce the existing categories of the context-aware recommendation models and then discuss existing work on interpreting contextual effects in the recommender systems.

To better understand the context-aware recommendation, we introduce the terminology in this domain as follows.

\begin{table}[ht!]
\centering
\caption{\label{tab:example}Contextual Ratings on Movies}
\begin{tabular}{|c|c|c|c|c|c|}\hline
\bfseries{User} & \bfseries{Item} & \bfseries{Rating} & \bfseries{Time} & \bfseries{Location} & \bfseries{Companion} \\ \hline

U1 & T1 & 3 & weekend & home & alone \\ \hline
U1 & T1 & 5 & weekend & cinema & girlfriend \\ \hline
U1 & T1 & ? & weekday & home & family \\ \hline
\end{tabular}
\end{table}

Assume there are one user $U1$, one item $T1$, and three contextual dimensions -- Time (weekend or weekday), Location (at home or cinema) and Companion (alone, girlfriend, family) as shown in the table above. In the following discussion, we use~\textit{context dimension} to denote the contextual variable, e.g. ``Location". The term~\textit{context condition} refers to a specific value in a dimension, e.g. ``home" and ``cinema" are two contextual conditions for ``Location". A \textit{context} or \textit{context situation} is, therefore, a set of contextual conditions, e.g. \{\textit{weekend, home, family}\}.

\subsection{Context-aware Recommendation Models}
\begin{figure}[ht!]
\centering
\includegraphics[scale=0.3]{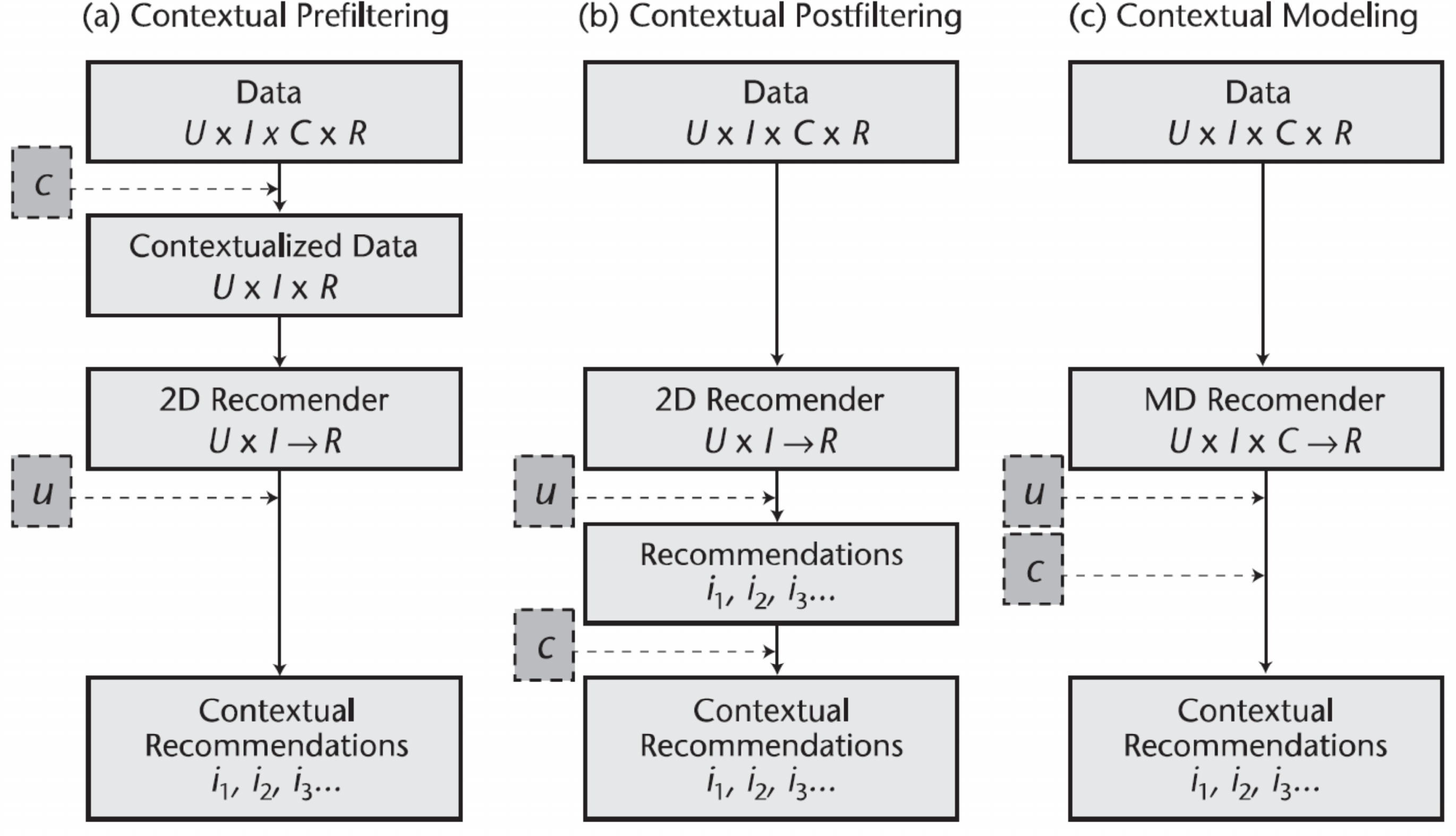}
\caption{Strategies for Context-aware Recommendation~\cite{carsmobasher}}
\label{fig:cars}
\end{figure}

Context can be applied in recommendation using three basic strategies: pre-filtering, post-filtering and contextual modeling~\cite{carsmobasher,adomavicius2011context}. The first two strategies rely on either filtering profiles or filtering the recommendations, but they still use standard two-dimensional recommendation algorithms in the modeling process. By contrast, in contextual modeling, the predictive models are learned by using the multidimensional rating data. These scenarios are depicted in Figure~\ref{fig:cars}~\cite{carsmobasher}.

As the name would suggest, pre-filtering techniques use the contextual information to remove profiles or parts of profiles from consideration in the recommendation process. For example, context-aware splitting approaches~\cite{baltrunas2009context,zheng2014splitting} use context as filter to pre-select rating profiles and then apply the recommendation algorithms only with profiles contain ratings in matching contexts.  Post-filtering techniques~\cite{panniello2009experimental,ramirez2014post} utilize contexts to filter or re-rank the list of the recommendations.

These filtering-based methods, including both pre-filtering and post-filtering, are straightforward and easy to be interpreted. However, by using context information as filters, it usually introduces the sparsity problem, and even the cold-start context problems. For example, to recommend a list of movies for a user within the context situation \{cinema, at weekend, with kids\}, we need rich rating profiles that the ratings were given in the same or similar contexts. Otherwise, the recommendations by the filtering-based approaches may be not that reliable.

By contrast, contextual modeling approaches are usually the machine learning based algorithms which are able to alleviate the sparsity problems and produce better context-aware recommendations than the ones by the filtering-based methods. These models directly incorporate context information as parts of the predictive functions, and contexts are no longer used as filters in the recommendation process. Tensor factorization~\cite{karatzoglou2010multiverse}, context-aware matrix factorization~\cite{baltrunas2011matrix} and contextual sparse linear modeling~\cite{zheng2014cslim} are the examples of the most effective contextual modeling algorithms in the recommender systems. However, it may be difficult to interpret the models in order to understand why and how contexts play an important role in the recommendation process.


\subsection{Interpretations By Contextual Filtering}
Due to the difficulty of interpreting the contextual modeling approaches, most of the existing work focus on the interpretations by the contextual filtering methods, especially the pre-filtering approaches. For example, our previous work~\cite{zheng2013emotions} views emotional states as the contexts, and utilizes the feature selection process in differential context relaxation (DCR)~\cite{zheng2012dcr} and the feature weighting in differential context weighting (DCW)~\cite{zheng2013dcw} to find out and interpret which emotional variables are crucial in different recommendation components or stages. DCR and DCW are two hybrid models of the contextual filtering approaches. Codina, et al.~\cite{codina2013exploiting} develop a distributional semantic pre-filtering context-aware recommendation algorithm which is able to calculate the similarity between two contexts. The context similarity, as a result, can tell why and which rating profiles are helpful in the recommendation process. It is able to alleviate the sparsity problems, since we can select similar rating profiles to predict a user's rating by given a context, and we no longer require an exact matching of the context information. As far as we know, there are no existing work that discuss the interpretations by the contextual modeling approaches, where we will explore this work in the following sections.

\section{Interpretation By Contextual Modeling}
In this paper, we are not going to develop new interpretable context-aware recommendation models. Instead, we focus on the existing contextual modeling approaches, reshape the structure of algorithms so that we can view these algorithms from another perspective and interpret the contextual effects by these approaches.

In this section, we introduce the dependent and independent contextual modeling approaches in our own way, and discuss the capability of the interpretations after we reshaping the structure of these algorithms.

\begin{figure*}[ht!]
\centering
\includegraphics[scale=0.8]{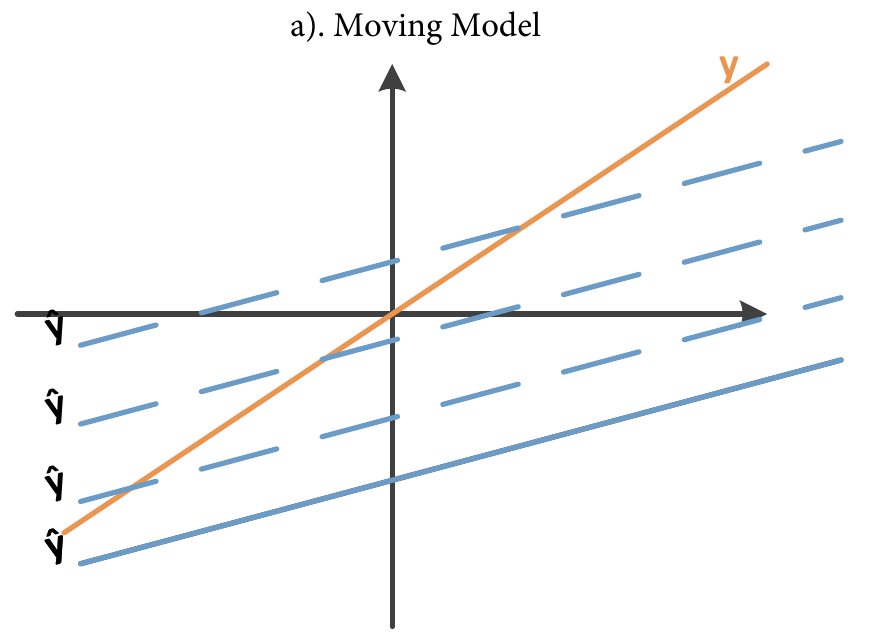}
\includegraphics[scale=0.8]{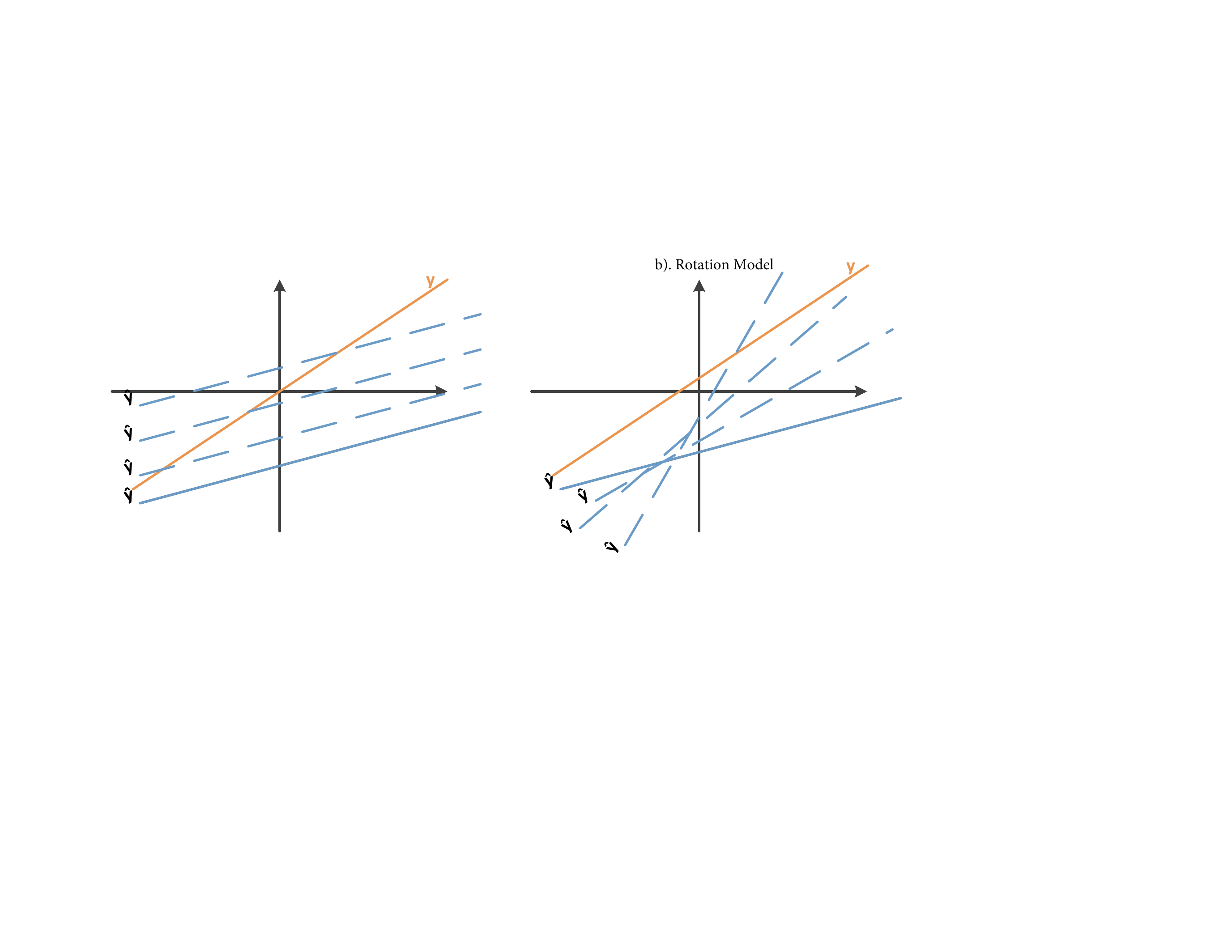}
\caption{Two Strategies In Linear Optimization}
\label{fig:linear}
\end{figure*}

\subsection{Example: A Linear Model}
First of all, we use the simple linear regression model (i.e., $ y = ax + b$) for example to explain two strategies in the optimization.

The visualization of the optimizations can be described by Figure~\ref{fig:linear}, where $y$ is the truth and $\hat y$ represents the estimated linear model. The optimization goal in the linear regression is to minimize the squared errors. Figure~\ref{fig:linear} a) is named as a moving model, since the model tries to vary the values of $b$ only in order to minimize the squared errors. Figure~\ref{fig:linear} b) is a rotation model, while this model varies the value of $a$ only in the optimization. Of course, the third strategy could be the one that combines the moving model and the rotation model, which is the common optimization in the linear regression model. In this paper, we only focus on the moving and rotation models, since they are simple and straightforward. We will explore the combination of these models in our future work.

\subsection{Dependent Contextual Modeling}
Among the contextual modeling approaches, dependent contextual model is the model that exploits the dependency or correlations among users, items and contexts. There are two existing categories of the algorithms -- deviation-based contextual modeling and similarity-based contextual modeling which we introduce individually as follows.

\subsubsection{Deviation-Based Models}
The deviation-based contextual modeling is a learning algorithm that minimizing the squared rating prediction errors by learning the rating deviations between two context situations. One example can be shown by Table~\ref{tab:dev}.

\begin{table}[ht!]
\centering
\caption{Example of Rating Deviations}
\label{tab:dev}
\begin{tabular}{|c|c|c|}
\hline
\textbf{Context} & \textbf{D1: Time} & \textbf{D2: Location} \\ \hline
$c_1$               & Weekend           & Home                  \\
$c_2$              & Weekday           & Cinema                \\ \hline
Dev(Di)          & 0.5               & -0.1                  \\ \hline
\end{tabular}
\end{table}

\begin{figure*}[ht!]
\centering
\includegraphics[scale=0.8]{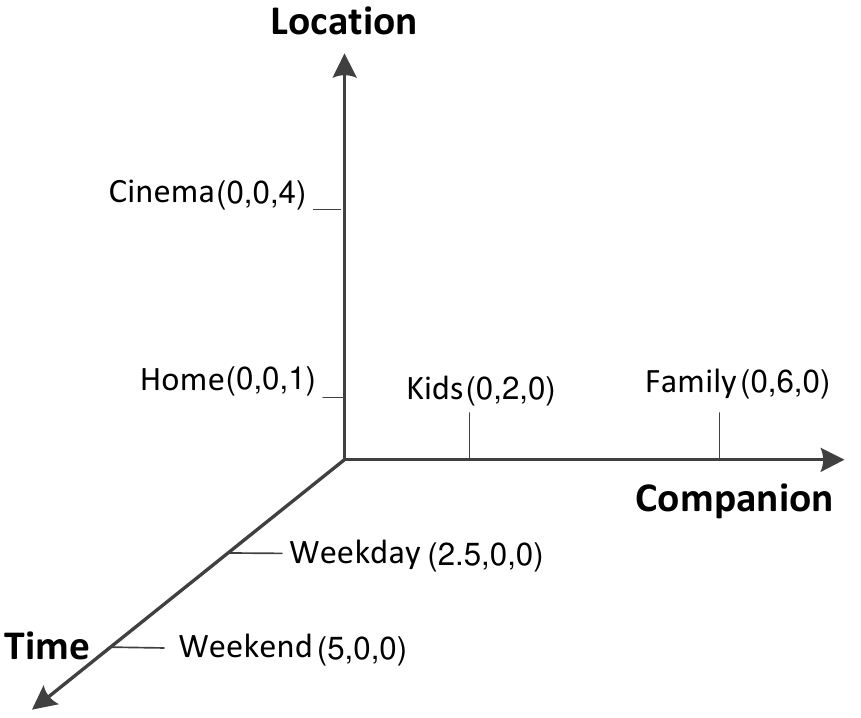}
\includegraphics[scale=0.8]{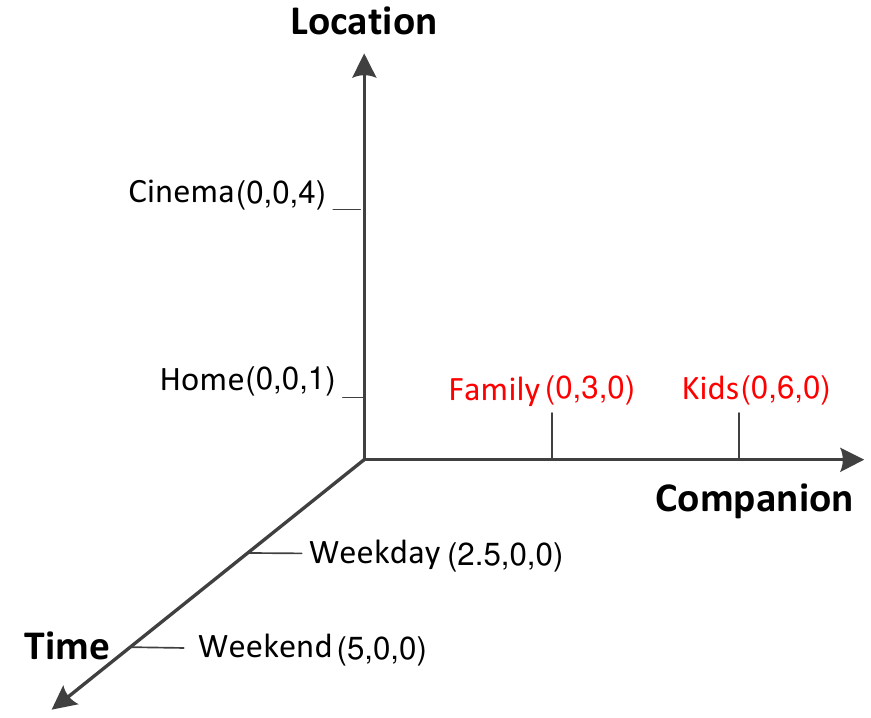}
\caption{Example of the Multidimensional Context Similarity}
\label{fig:mcs}
\end{figure*}

Assume there are two context dimensions: Time and Location. Each context situation is constructed by the context conditions in these two variables. There are two contexts $c_1$:\{Weekend, Home\} and $c_2$:\{Weekday, Cinema\}. The last row in Table~\ref{tab:dev} tells the rating deviation from $c_1$ to $c_2$ in each context variable. For example, Dev(D1) represents the rating deviation in the variable "Time" from $c_1$ to $c_2$. More specifically, it indicates that a user's rating in weekday is generally higher than his or her rating at weekend by 0.5. Accordingly, Dev(D2) is -0.1, which tells that a user's rating in cinema is generally lower than his or her rating at home by 0.1.

Therefore, we can predict a user's rating on a specific item within contexts $c_2$ if we know his or her rating on the same item within contexts $c_1$. For example, if user $u$ rated item $t$ in context $c_1$ as a four star, his or her rating on $t$ in context $c_2$ can be simply estimated as the four star plus the aggregated rating deviations in each context variable. Namely, the predicted rating will be 4.4 (i.e., 4 + 0.5 - 0.1).

Theoretically, we are able to learn the rating deviations between every two context conditions in a same context variable. However, it may introduce sparsity problems if there are many context conditions in a single context variable. A simple solution to alleviate this problem is to set a baseline. Take Table~\ref{tab:dev2} for example, we introduce a special context situation $c_0$, where the context conditions in all the context variables are "N/A" (i.e., not available). The ratings in $c_0$ can be interpreted as a user's ratings on the items without considering contextual situations.

\begin{table}[ht!]
\centering
\caption{Example of Rating Deviations}
\label{tab:dev2}
\begin{tabular}{|c|c|c|}
\hline
\textbf{Context} & \textbf{D1: Time} & \textbf{D2: Location} \\ \hline
$c_0$               & N/A          & N/A                 \\
$c_2$              & Weekday           & Cinema                \\ \hline
Dev(Di)          & 0.5               & -0.1                  \\ \hline
\end{tabular}
\end{table}

Therefore, the predictive function for user's rating on an item within a context can be described as Equation~\ref{eq:dev}.
\begin{equation}
\label{eq:dev}
F(u, t, c) = P(u, t) + \sum_1^N Dev(D_i)
\end{equation}

where $F(u, t, c)$ is the prediction function to estimate user $u$'s rating on item $t$ within context $c$. $P(u, t)$ is the predicted rating given by $u$ on $t$ without considering any context situations. $Dev(D_i)$ tells the rating deviation at the $i^{th}$ context variable from $c_0$ to $c$, where $N$ is the number of context dimensions in the data set.

$P(u, t)$, as the predicted rating given by $u$ on $t$, can be replaced by any predictive function in the traditional recommendation algorithms. For example, it could be the prediction function in user-based collaborative filtering, or the function by matrix factorization. What we are going to learn are the rating deviations in each context variable from $c_0$ to $c$, i.e., the rating deviation between two context conditions in each context variable.

In Equation~\ref{eq:dev}, we simply assume the $Dev(D_i)$ is the same for all the users and the items. A finer-grained model may assume $Dev(D_i)$ may vary from users to users, from items to items. For example, a user-specific model could be described by Equation~\ref{eq:dev2}, where we assign a $Dev(D_i)$ to each user by assuming that different users may have personalized values in $Dev(D_i)$. According, an item-specific model can be developed too.

\begin{equation}
\label{eq:dev2}
F(u, t, c) = P(u, t) + \sum_1^N Dev(D_i, u)
\end{equation}

The introductions above give a high-level picture of how the rating deviations in different contexts can be incorporated into the recommendation model. Context-aware matrix factorization (CAMF)~\cite{baltrunas2011matrix} is the first attempt as the deviation-based contextual modeling approach, where it replaces the $P(u, t)$ by the predictive function in matrix factorization. Deviation-based contextual sparse linear method (CSLIM)~\cite{zheng2014cslim} is another example which utilizes the prediction function in sparse linear method~\cite{ning2011slim} as the component $P(u, t)$.

The deviation-based contextual modeling is similar to the moving model described in Figure~\ref{fig:linear}, where the component of the aggregation of rating deviations (such as $\sum_1^N Dev(D_i)$) is equivalent to the $b$ in the linear regression model.

\subsubsection{Similarity-Based Models}
By contrast, similarity-based contextual modeling tries to learn the similarity between two context situations.

\begin{table}[ht!]
\centering
\caption{Example of Context Similarity}
\label{tab:sim}
\begin{tabular}{|c|c|c|}
\hline
\textbf{Context} & \textbf{D1: Time} & \textbf{D2: Location} \\ \hline
$c_0$               & N/A          & N/A                 \\
$c_2$              & Weekday           & Cinema                \\ \hline
Sim(Di)          & 0.5               & 0.1                  \\ \hline
\end{tabular}
\end{table}

Table~\ref{tab:sim} gives an example of context similarity. The table is similar to the one that is used to represent rating deviations in contexts. The last row in Table~\ref{tab:sim} tells the similarity of the contexts between $c_0$ and $c_2$ in each context variable. In fact, the term "context similarity" or "similarity of contexts" actually refers to the similarity of a user's rating behavior in two contexts.

Therefore, the predicted rating or ranking score can be described as:

\begin{equation}
\label{eq:sim}
F(u, t, c) = P(u, t) \times Sim(c_0, c)
\end{equation}

Again, any predictive function in the traditional recommender systems can be used to replace the $P(u, t)$. The challenge becomes how to measure the similarity between $c_0$ and $c$.

We~\cite{zheng2015umap,zheng2015simcars} propose three methods to represent the similarity of contexts. The independent context similarity (ICS) assumes the similarity between two contexts equals to the multiplication of the similarity between two context conditions in each context variable. In this case, the model will learn the similarity between every two context conditions in the same context variable. The latent context similarity (LCS) is an improved method based on the ICS, where each context condition is represented by a latent vector, and therefore the similarity between two context conditions can be estimated by the dot product of the two corresponding vectors. LCS is used to alleviate the cold-start context problem in ICS.

The multidimensional context similarity (MCS) is the most effective but also the complicated one. An example of the visualization can be shown by Figure~\ref{fig:mcs}. In MCS, each context variable is depicted by a dimension or an axis in the multidimensional space. Each context condition will be assigned a real number value so that they can be placed in specific positions. As a result, a context situation, such as \{Weekend, Family, Home\}, can be represented as a point in the space. The dissimilarity of two context situations becomes the distance between two points. In MCS, the model will learn the position of the context conditions. For example, in Figure~\ref{fig:mcs}, we change the positions of "Family" and "Kids", which results in difference distance values between two contexts, since the position of the points will be changed if the values for the two context conditions, "Family" and "Kids", are updated.

Anyway, the similarity-based contextual modeling is able to learn the similarity of contexts. Note that this approach is different from the semantic pre-filtering algorithm. As mentioned previously, Codina, et al.~\cite{codina2013exploiting} develop a distributional semantic pre-filtering context-aware recommendation algorithm which is able to calculate the similarity between two contexts. In their approach, the context similarity is calculated based on a formula, while the similarity of contexts is learned by minimizing squared errors in the learning process.

The similarity-based contextual modeling is similar to the rotation model we introduced previously, where we try to vary the value of the $a$ in the linear regression model. Of course, it is able to combine the moving and rotation model. Similarly, we may also combine the deviation-based and similarity-based contextual modeling, which we will explore in our future work.

\subsection{Independent Contextual Modeling}
Independent contextual modeling doesn't explicitly make assumptions of the dependency or correlations among users, items and contexts. These models assume the user, item and context dimensions are independent and explore the interactions among these three dimensions. One example is the tensor factorization (TF)~\cite{karatzoglou2010multiverse} that was applied in the context-aware recommendations. The optimization in TF can be realized by either Tucker decomposition or Canonical Polyadic (CP) decomposition. We choose CP decomposition in this paper, since each context condition can be represented by a vector. Therefore, we can continue to calculate the similarity of contexts by using the latent context similarity (LCS)~\cite{zheng2015simcars} and the learned vectors in TF.

\subsection{Capability of Interpretations}
We interpret the dependent and independent contextual modeling approaches in our own way in the previous sections. We summarize the capability of interpretations of these contextual modeling methods as follows:

\begin{itemize}
    \item In dependent contextual modeling, the models explicitly make assumptions about the dependency or correlations among users, items and contexts. For example, the deviation-based models can learn the rating deviations in different contexts. These deviations could be user-specific or item-specific ones. By contrast, independent contextual modeling may be more difficult to interpret the contextual effects in the recommendation process.
    \item The deviation-based contextual modeling can tell the rating deviations between two contexts. For example, a user's rating in one context condition is higher or lower than another context condition. It interprets why a user gives a higher rating in one context situation than another situation, even if the user watches the same movie or listens to a same music track.
    \item The similarity-based contextual modeling learns the similarity between two contexts. It is pretty useful since we do not need to perform exact matching in two context situations, but we can tell how similar the user ratings will be if they are going to leave ratings within two contexts.
    \item It is much more difficult to interpret the independent context modeling. By using CP decomposition in the TF, we can interpret the similarity of contexts by using the learned vectors in TF and latent context similarity.
\end{itemize}

\section{Experiment and Results}
In this paper, we decide to focus on interpreting the contextual effects or the contextual modeling by comparison of the context similarities, and ignore the rating deviations in different contexts. The underlying reasons can be listed as follows:

\begin{itemize}
    \item We can obtain the rating deviations in different contexts, but it is difficult to evaluate whether they are true or not. By contrast, we can examine the context similarity by common sense. For example, given a context situation, we can utilize the similarity-based contextual modeling and the TF approach to retrieve the top-$N$ similar contexts, and compare the lists by our common sense.
    \item The TF approach can only interpret the context similarities, if we would like to add the independent contextual modeling in the experimental comparison.
    \item We can also compare the quality of context similarity between contextual modeling and the contextual pre-filtering approach that was developed by odina, et al.~\cite{codina2013exploiting}.
\end{itemize}

\begin{figure*}[ht!]
\centering
\includegraphics[scale=0.7]{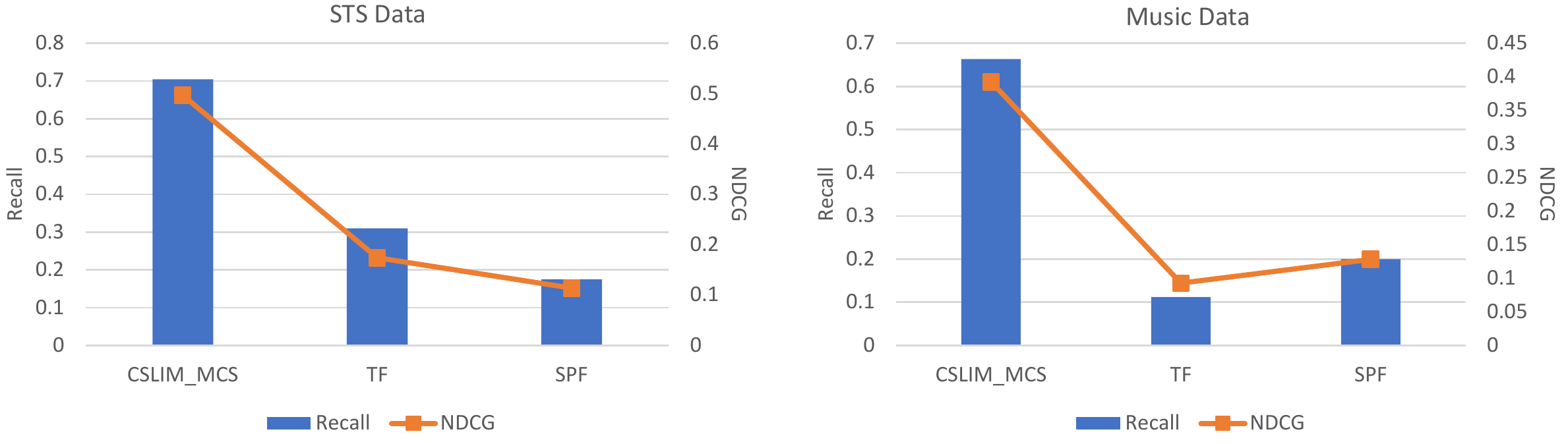}
\caption{Quality of Top-10 Context-aware Recommendations}
\label{fig:top10}
\end{figure*}

\begin{table*}[ht!]
\centering
\caption{Top-5 Similar Context Situations}
\label{tab:top5}
\begin{tabular}{|c|c|c|c|}
\hline
\multicolumn{4}{|c|}{\textbf{STS Data (temperature: cold)}}                                                                                                                                                                                                                            \\ \hline
\textbf{} & \textbf{CSLIM\_MCS}                                                                                   & \textbf{TF}                                                                  & \textbf{SPF}                                                                        \\ \hline
1         & \begin{tabular}[c]{@{}c@{}}daytime: morning\\ temperature: cold\end{tabular}                          & \begin{tabular}[c]{@{}c@{}}weather: cloudy\\ temperature: cold\end{tabular}  & \begin{tabular}[c]{@{}c@{}}daytime: morning\\ temperature: cold\end{tabular}        \\ \hline
2         & \begin{tabular}[c]{@{}c@{}}season: winter\\ temperature: cold\end{tabular}                            & \begin{tabular}[c]{@{}c@{}}daytime: morning\\ temperature: cold\end{tabular} & \begin{tabular}[c]{@{}c@{}}weather: cloudy\\ temperature: cold\end{tabular}         \\ \hline
3         & \begin{tabular}[c]{@{}c@{}}distance: near by\\ temperature: cold\end{tabular}                         & \begin{tabular}[c]{@{}c@{}}season: winter\\ temperature: cold\end{tabular}   & \begin{tabular}[c]{@{}c@{}}time available: one day\\ temperature: cold\end{tabular} \\ \hline
4         & \begin{tabular}[c]{@{}c@{}}daytime: night\\ time available: half day\\ temperature: cold\end{tabular} & \begin{tabular}[c]{@{}c@{}}daytime: night\\ weather: cloudy\end{tabular}     & season: winter                                                                      \\ \hline
5         & \begin{tabular}[c]{@{}c@{}}weather: cloudy\\ temperature: cold\end{tabular}                           & season: winter                                                               & mood: lazy                                                                          \\ \hline\hline
\multicolumn{4}{|c|}{\textbf{Music Data (landscape: country side)}}                                                                                                                                                                                                                    \\ \hline
\textbf{} & \textbf{CSLIM\_MCS}                                                                                   & \textbf{TF}                                                                  & \textbf{SPF}                                                                        \\ \hline
1         & landscape: mountains                                                                                  & naturalphenomena: afternoon                                                  & naturalphenomena: afternoon                                                         \\ \hline
2         & roadtype: serpentine                                                                                  & naturalphenomena: morning                                                    & landscape: mountains                                                                \\ \hline
3         & roadtype: city                                                                                        & roadtype: city                                                               & naturalphenomena: morning                                                           \\ \hline
4         & naturalphenomena: afternoon                                                                           & landscape: mountains                                                         & roadtype: city                                                                      \\ \hline
5         & trafficconditions: free road                                                                          & roadtype: serpentine                                                         & naturalphenomena: day time                                                          \\ \hline
\end{tabular}
\end{table*}

\subsection{Experimental Setting}
In the domain of context-aware recommendation, there are very limited number of available data sets for research. One of the reasons is that the context acquisition is difficult. Therefore, most data were collected from surveys, which results in either small or sparse data set. Another reason is that the context information is usually related to user privacy problem. It is difficult to obtain the large context-aware data set from industry.

There are a list of available context-aware data sets\footnote{https://github.com/irecsys/CARSKit/tree/master/context-aware\_data\_sets}, where we select the South Tyrol Suggests (STS) and the music data, since they have many more context dimensions and conditions. More specifically,
\begin{itemize}
    \item the STS data~\cite{braunhofer2013context} was collected from a mobile app which provides
context-aware suggestions for attractions, events, public services, restaurants, and much more
for South Tyrol. There are 14 contextual dimensions in total, such as budget, companion,
daytime, mood, season, weather, etc. The total number of context conditions from all of the 14 dimensions is 53. There are 2354 ratings (scale 1-5) given by 325 users on 249 items within different context situations.
    \item the music data~\cite{baltrunas2011incarmusic} was collected from InCarMusic which is an Android mobile application
offering music recommendations to the passengers of a car. Users are requested to enter ratings
for some items using a web application. There are 8 context dimensions included in
the data: driving style, road type, landscape, sleepiness, traffic conditions, mood, weather,
natural phenomena. There are 26 context conditions in total. 3251 ratings (scale 1-5) were given by 42 users on 139 items within different context situations.
\end{itemize}

\subsection{Evaluation Protocol}
Our add the following models in the comparison: tensor factorization (TF)~\cite{karatzoglou2010multiverse}, similarity-based contextual sparse linear method using multidimensional context similarity (CSLIM\_MCS)~\cite{zheng2015simcars} and the semantic pre-filtering algorithm (SPF)~\cite{codina2013exploiting}. TF is the representative of independent contextual modeling, while CSLIM\_MCS is demonstrated as the best performing dependent contextual modeling approach which utilizes context similarity. SPF is our baseline which is a pre-filtering algorithm that utilizes the context similarity.

First of all, we examine the quality of top-10 context-aware recommendations by these algorithms. We select recall and normalized discounted cumulative gain (NDCG) as the evaluation metrics. After that, we retrieve the top-5 similar context situations by given a context, and compare the list of ranked contextual situations to see which approach is better. We use CARSKit~\cite{zheng2015carskit} which is an open-source context-aware recommendation library in our experiments. All of these evaluations are based on 5-fold cross-validation, since the data is relatively small.

\subsection{Results and Findings}
The quality of the top-10 context-aware recommendations can be described by Figure~\ref{fig:top10}, where the bars present the results in precision and the curves describe the results in NDCG which tells the quality of the rankings. Apparently, CSLIM\_MCS is the best performing model which obtains the highest precision and NDCG in these two data sets. TF performs better than the SPF approach in the STS data, but SPF is the better one in the music data. Dependent contextual modeling is usually better than the independent modeling, since there are always dependency among users, items, contexts in the data, especially when it comes to a data set that has many more context dimensions. It is not surprising that TF fails to outperform CSLIM\_MCS in these two data sets.

Afterwards, we want to compare whether these models can retrieve high-quality similar contexts. In the STS data, we pick up the "temperature: cold" as the target context situation. We run CSLIM\_MCS and SPF algorithms on the STS data and try to output the top-5 similar contextual situations to this target context. In terms of the TF approach, we utilize the latent context similarity (LCS)~\cite{zheng2015simcars} based on the learned vectors in TF to retrieve the top-5 similar contexts. We choose "landscape: country side" as the target context and apply the same process in the music data.

The results are described by Table~\ref{tab:top5}. Note that, there is only one context condition for each rating profile in the music data, while the context situation is a combination of context conditions in the STS data. Based on the results in Table~\ref{tab:top5}, we can observe that the CSLIM\_MCS approach can retrieve more contextual situations with the combinations of "temperature: cold" in the STS data, while the TF and SPF retrieve context conditions in other dimensions, such as season, daytime, mood, without the pair of "temperature: cold". It may be useful to explain why CSLIM\_MCS outperforms other approaches, since it is better to learn the similarity of contexts.

Similar patterns can be observed in the music data. For example, CSLIM\_MCS can retrieve more relevant context situations such as the conditions in landscape and roadtype dimensions, while TF and SPF retrieve more situations in the naturalphenomena dimension.

Based on these comparisons, we can tell CSLIM\_MCS is the best performing context-aware recommendation algorithm for these two data sets. According to the retrieved top-5 similar contexts, we can observe that CSLIM\_MCS is also able to better learn and retrieve more relevant context situations by given a target context. By this way, we are able to understand why one contextual modeling is better than other algorithms by the comparison and interpretations.

\section{Conclusions and Future Work}
In this paper, we focus on the capability of interpretations by the contextual modeling approaches in the recommender systems. We reshape or re-explain the structure of the independent and dependent contextual modeling, and exploit the interpretations of contextual effects based on the similarity of the contexts. Our experimental results on the STS and music data can tell that CSLIM\_MCS is the best performing context-aware recommendation algorithms for these two data sets, since it is able to better learn the similarity of the contexts.

In this paper, we did not evaluate the capability of interpretations by the deviation-based contextual modeling. In our future work, we will seek appropriate ways to evaluate the quality of rating deviations in different contexts, and also try to figure out a way to compare the interpretations by the deviation-based and similarity-based contextual modeling.
